\documentclass[preprint,prl,amsmath,amssymb,superscriptaddress]{revtex4}

\usepackage{graphicx}
\usepackage{dcolumn}
\usepackage{bm}
\usepackage{ulem}
\usepackage{color}
\usepackage{amsmath}
\usepackage[sort&compress]{natbib}

\begin{document}

\title{Resonant interactions between a Mollow triplet sideband of a quantum dot and a strongly coupled cavity.}

\author{Hyochul Kim}
\affiliation{Department of Electrical and Computer Engineering, IREAP,
and Joint Quantum Institute, University of Maryland, College Park, Maryland 20742, USA}

\author{Thomas C. Shen}
\affiliation{Department of Electrical and Computer Engineering, IREAP,
and Joint Quantum Institute, University of Maryland, College Park, Maryland 20742, USA}

\author{Kaushik Roy-Choudhury}
\affiliation{Department of Electrical and Computer Engineering, IREAP,
and Joint Quantum Institute, University of Maryland, College Park, Maryland 20742, USA}

\author{Glenn S. Solomon}
\affiliation{Joint Quantum Institute, National Institute of Standards and Technology, and
University of Maryland, Gaithersburg, Maryland 20899, USA}

\author{Edo Waks} \email{edowaks@umd.edu}
\affiliation{Department of Electrical and Computer Engineering, IREAP,
and Joint Quantum Institute, University of Maryland, College Park, Maryland 20742, USA}

\begin{abstract}
We experimentally demonstrate that the Mollow triplet sidebands of a
quantum dot strongly coupled to a cavity exhibit anomalous power induced
broadening and enhanced emission when one sideband is tuned over the cavity
frequency. We observe a nonlinear increase of the sideband linewidth with
excitation power when the Rabi frequency exceeds the detuning between the
quantum dot and the cavity, consistent with a recent theoretical model that accounts for
acoustic phonon-induced processes between the exciton and the cavity.
In addition, the sideband tuned to the cavity shows strong resonant
emission enhancement.
\end{abstract}

\maketitle

Resonance fluorescence of atomic systems is important for studying
light-matter interactions, and also plays a crucial role in quantum
information processing. For example, it can be used to perform
non-destructive quantum state
readout~\cite{SauterPRL86,BergquistPRL86,VamivakasNature10} and generate
single photons~\cite{KimblePRL77,MatthiesenPRL12}. In these cases, the laser
power is typically well below the saturation power of the quantum dot
exciton. Under strong pump excitation, however, the resonance fluorescence
emission becomes highly nonlinear and exhibits a Mollow triplet
spectrum~\cite{Mollow69}.

The Mollow triplet spectrum has been observed with atoms~\cite{SchudaJPB74}
as well as solid-state quantum emitters such as quantum
dots~\cite{MullerPRL07,XuSci07,MullerPRL08,VamivakasNPhys09,FlaggNPhys09,AtesPRL09}.
Unlike single isolated atoms, quantum dots are coupled to a broad continuum
of phonon states in a semiconductor matrix that significantly affects the
resonance fluorescence properties~\cite{forstnerPRL03,RamsayPRL10}. This
phonon mediated interaction is particularly strong when the quantum dot is
coupled to a cavity. Cavity enhancement of phonon effects has been
investigated experimentally and theoretically in the small Rabi frequency
regime where the Mollow triplet sidebands are highly detuned from the
cavity~\cite{UlrichPRL11,RoyPRL11,ArkaArxiv11,ArkaPRB11,HughPRX11}. In this
regime, phonons cause power broadening of the sidebands, whose linewidth
linearly increases with the excitation power~\cite{UlrichPRL11,RoyPRL11}. In
the large Rabi frequency regime, however, recent theoretical work has
predicted \textit{anomalous} broadening of the sideband
emission~\cite{RoyArx12} where the linewidth becomes a highly nonlinear
function of pump power. In particular, the rate of increase of the sideband
linewidth is expected to significantly change when the emission frequency of
one of the sidebands crosses the cavity resonance, which is the transition
point between the small Rabi frequency and large Rabi frequency regimes. The
large Rabi frequency regime, however, has yet to be experimentally attained
and anomalous power broadening has not been demonstrated.

In this paper, we experimentally demonstrate that Mollow triplet sidebands of
a quantum dot strongly coupled to a cavity exhibit anomalous power broadening
and large emission enhancement when tuned over the cavity frequency. We
obtain a Rabi frequency exceeding 100~GHz, which enables us to tune the
sideband emission over the entire resonant spectrum of the cavity. We show
that the emission intensity of the cavity-resonant sideband is strongly
enhanced relative to the intensity of the detuned sideband by as much as a
factor of 6.  We also observe an anomalous power induced
broadening when the sideband crosses the cavity resonance, consistent with
recent theoretical predictions~\cite{RoyArx12}. We compare our experimental
results to numerical simulations based on an effective phonon master equation,
derived from a full polaron model~\cite{HughPRX11}, which show excellent agreement.

The system we study is an InAs quantum dot strongly coupled to a GaAs
photonic crystal three-hole defect cavity~\cite{AkahaneOptExp05}.
Fig.~1a shows a scanning electron microscope image of a fabricated cavity.
The sample consists of 160~nm GaAs membrane with a single layer of InAs
quantum dots (density of $10-50/\mu$m$^2$) at the center. The GaAs membrane
is grown on top of a 1~$\mu$m AlGaAs sacrificial layer. We fabricate photonic
crystal cavities using electron beam lithography and chlorine based dry
etching, followed by a wet etch of the sacrificial layer to create a
membrane.

Fig.~1b shows the measurement setup. Sample excitation and collection are
performed with a confocal microscope using an objective lens with numerical
aperture of 0.68. We excite the sample using either a Ti:Sapphire laser
emitting at 780~nm for photoluminescence measurements, or with a narrow
linewidth ($<300$~kHz) tunable diode for resonance fluorescence measurements.
The collected signal is measured using a grating spectrometer with a
resolution of 7~GHz and detected by a charged coupled device (CCD). A
cross-polarization setup rejects the component of the excitation laser that
does not couple to the cavity mode. For high-resolution spectral
measurements, we use a tunable fiber Fabry-Perot filter with 0.9~GHz
bandwidth in front of the spectrometer.

We first characterize the device through photoluminescence. Fig.~1c shows the
measured cavity spectrum as a function of the sample temperature. The
spectrum shows two resonances, one which corresponds to the direct cavity
emission and a second due to a quantum dot.  As the sample temperature
increases, the quantum dot resonance red-shifts and exhibits a clear
anti-crossing with the cavity indicating that the system operates in the
strong coupling regime. A minimum energy separation ($\Delta \textrm{E}$) of
101~$\mu$eV (24.7~GHz) is observed on resonance at 18.5~K. When the quantum
dot is detuned from the cavity, the photoluminescence from the bare cavity
shows a quality factor of 9,100, corresponding to a cavity energy decay rate
of $\kappa/ 2\pi=36$~GHz. The coupling strength $g$ is calculated to be $g/2
\pi=15.3$~GHz using the relation $\Delta \textrm{E} = 2 \hbar
\sqrt{g^2-(\kappa/4)^2}$.

Fig.~1d shows the resonance fluorescence spectrum taken when the narrow band
diode laser is swept over the quantum dot emission wavelength at temperature
of 5~K. The diode laser power is set to 8.0~$\mu$W and the signal is measured
directly by the spectrometer. Using the cross-polarization setup and spatial
filtering with a single mode fiber, we reject a large fraction of the direct
laser scatter from the sample surface. The cavity mode is blue-detuned from
the quantum dot by 0.24~nm and lies outside the plotted wavelength range.
When the laser is tuned close to the quantum dot emission line (927.64~nm),
the spectrum exhibits a Mollow triplet.

Fig.~2a shows the resonance fluorescence spectrum as a function of
$\sqrt{P}$, where $P$ is the pump power measured before the focusing lens.
The $x$-axis is relative frequency $\omega'=\omega-\omega_{L}$, where $\omega$ is the
measured frequency and $\omega_{L}$ is a laser frequency. The detuning
between the quantum dot emission frequency ($\omega_x$) and cavity
($\omega_c$) is $\Delta_{cx}/2\pi=(\omega_c-\omega_x)/2 \pi=42$~GHz (0.12~nm),
and the
excitation laser is tuned to resonance with the quantum dot. At each power,
the Mollow triplet sidebands appear symmetrically on the longer and shorter
wavelength sides of the quantum dot emission and their spectral separation
increases linearly with $\sqrt{P}$. The Rabi frequency $\Omega$ is related to
the splitting between the two sidebands, denoted $\Delta\omega$, via the
relation $\Omega=1/2~\Delta\omega$. At the maximum pump power we achieve a
Rabi frequency exceeding $\Omega/2\pi=100$~GHz, which is much greater than
$\Delta_{cx}/2\pi$.  Thus we are able to drive the system both in the small Rabi
frequency regime ($\Omega<\Delta_{cx}$) and the large Rabi frequency regime
($\Omega>\Delta_{cx}$).

At a pump power of $13.4~\mu$W, the higher energy sideband is resonant with the cavity mode (black dashed
line). A clear enhancement of emission is observed under this condition,
resulting in a highly asymmetric spectrum where the higher energy sideband
shows a higher intensity than the lower energy sideband. At even higher pump power, the
higher energy sideband tunes beyond the cavity resonance and is
once again diminished. We note that at very high pump power the bare cavity
emission appears in the spectrum. This emission is attributed to phonon
induced non-resonant energy transfer of the quantum dot
excitation~\cite{AtesNPhoton09,HohenesterPRB09,WingerPRL09,EnglundPRL10,HughesPRB11}.

In Fig.~2b, we plot the emission intensity of the higher (blue open squares)
and lower (red full circles) energy sidebands  as a function of the Rabi
frequency. To determine the Rabi frequency and intensity, we fit the measured
spectrum at each laser power to four Lorentzians, one for each sideband, one
for the laser scatter, and one for the cavity which is excited by inelastic
scattering. When the higher energy sideband is within 10~GHz of the cavity
resonance, it becomes difficult to separate it from the cavity emission due
to inelastic scattering.  In this region, we interpolate the inelastic
scattering intensity using the closest data points outside the 10~GHz window.
Fig.~2b shows a clear resonant behavior where the sideband is
enhanced near cavity resonance, resulting in a large emission asymmetry. The
higher energy sideband is 6 times brighter than the lower energy sideband at
a pump power of $24.6~\mu$W ($\Omega/2\pi=57.8$~GHz).

Fig.~2c-d show the spectrum as a function of pump power where the detuning
between the cavity and quantum dot is increased to $\Delta_{cx}/2\pi=85$~GHz (0.24~nm) by gradual gas condensation that occurs naturally in the
vacuum chamber~\cite{MosorAPL05,StraufAPL06}. Here, the detuning is greater
than the maximum achievable Rabi frequency so the system remains in the small Rabi
frequency regime for all pump powers.  Because of the larger detuning, the
Mollow sideband does not cross the cavity resonant frequency and we do not
observe a resonance behavior. Instead, the sideband emission intensity
gradually increases as it tunes closer to resonance with the cavity.

In addition to the asymmetry and intensity increase at cavity resonance,
Fig.~2a shows indications of linewidth broadening.  The sideband linewidth is
predicted to exhibit an anomalous power broadening behavior in the large Rabi
frequency regime~\cite{RoyArx12}, where it becomes a highly nonlinear
function of pump power.  Due to the resolution limit of the spectrometer,
however, this broadening is difficult to resolve from the data in Fig.~2. In
order to improve the spectral resolution of the measurement system we place a
tunable fiber Fabry-Perot filter with 0.9~GHz bandwidth in front of the
spectrometer. Measurements are performed by tuning the filter and recording
the intensity, determined by integrating the signal over a 14~GHz spectral
window (corresponding 5 pixels of CCD) around the center frequency of the
Fabry-Perot mode. We measure the linewidth of the lower energy sideband since
it is always highly detuned from the cavity and therefore spectrally well
separated from the background photons created by non-resonant energy
transfer.

Fig.~3a shows the measured high-resolution spectrum of the lower energy
sideband for several different pump powers at a detuning of
$\Delta_{cx}/2\pi=42$~GHz. For each spectrum, the sideband linewidth
is determined by fitting the measured data with two-Lorentzian functions, one
representing the sideband peak while the other the direct laser signal. The
fit is shown as a solid red line in the figure. Fig.~3b shows the measured
full-width half-maximum linewidth of the Mollow sideband as a function of
$|\Omega/2\pi|^2$. Here $\Omega$ is determined by measuring the detuning of the
lower energy sideband from the laser, where the sideband center frequency is
obtained from the Lorentzian fit. In the small Rabi frequency regime ($\Omega< \Delta_{cx}$)
we observe a linear increase in the sideband linewidth as a
function of pump power (proportional to $|\Omega|^2$). At a pump power of
$|\Omega/2\pi|^2=2040~$GHz$^2$ (denoted by the dashed
vertical line), the higher energy sideband becomes resonant with the cavity
and the system transitions to the large Rabi frequency regime ($\Omega>\Delta_{cx}$).
At this point the linewidth exhibits an anomalous broadening
behavior where it becomes a highly nonlinear function of pump power.  In the
large Rabi frequency regime the linewidth is largely insensitive to the pump
power. We observe the transition between these two behaviors precisely at the
point where the higher energy sideband crosses the cavity mode. On the other
hand, Fig.~3c shows the lower energy sideband linewidths obtained for a
larger quantum dot-cavity detuning ($\Delta_{cx}/2\pi=85$~GHz) where
the system remains in the small Rabi frequency regime. Here the linewidth
shows nearly linear increase over the same range of excitation power.

To gain further insight into the mechanism for the anomalous power broadening
behavior we perform numerical simulations of the master equation $\dot{\rho}
=-i/\hbar~[\mathbf{H},\rho]+\mathbf{L}\rho,$ where $\rho$ is the density
matrix of the system. The system Hamiltonian is given by
\begin{equation}
\mathbf{H}=\hbar\Delta_c \mathbf{a^{\dag}a}+\hbar\Delta_x \mathbf{\sigma_z}/2+
\hbar g (\mathbf{\sigma_+ a}+\mathbf{a^{\dag} \sigma_-} )+ \hbar\sqrt{\kappa}J (\mathbf{a}+\mathbf{a^{\dag}}).
\end{equation}
In Eq.~1, $\Delta_c=\omega_c-\omega_L$ and $\Delta_x=\omega_x-\omega_L$.
In addition,
$\mathbf{\sigma_z}$ is the population difference operator between the excited
and ground state of the quantum dot, $\mathbf{\sigma_-}$
($\mathbf{\sigma_+}$) represents the dipole lowering (raising) operator for the
quantum dot, $\mathbf{a}$ ($\mathbf{a}^{\dag}$) is the cavity photon
annihilation (creation) operator, and $J=\sqrt{\eta P/\hbar\omega}$ is the
driving field amplitude. The Liouvillian superoperator $\mathbf{L}$ accounts
for all non-unitary Markovian processes including cavity and quantum dot
damping, pure dephasing, and phonon mediated energy transfer.  This operator can
be written as
\begin{equation}
\begin{split}
\mathbf{L}=
& \gamma \mathcal{D} (\sigma_-)+\kappa \mathcal{D} (a)+\gamma_d \mathcal{D} (\sigma_+\sigma_-)
\\
&  +\gamma_{ph}^{a^{\dag} \sigma_-} \mathcal{D} (a^{\dag} \sigma_-)+\gamma_{ph}^{a \sigma_+} \mathcal{D} (a \sigma_+),
\end{split}
\end{equation}
where $\mathcal{D} (\mathcal{C})\rho=\mathcal{C}\rho\mathcal{C}^{\dag}-1/2
\mathcal{C}^{\dag}\mathcal{C}\rho-1/2 \rho\mathcal{C}^{\dag}\mathcal{C}$ is a
general Linblad operator form for the collapse operator $\mathcal{C}$.  In
Eq.~2, $\gamma$ is the quantum dot spontaneous emission rate, $\kappa$ is the
cavity energy decay rate, and $\gamma_d$ is the quantum dot pure dephasing
rate.  To account for the phonon mediated dephasing, we adopt the formalism
of Ref.~\cite{HughPRX11} and include the last two Linblad terms where we
define $\gamma_{ph}^{a^{\dag} \sigma_-}$ and $\gamma_{ph}^{a \sigma_+}$ as
the phonon mediated dephasing rates;
physically these processes describe the destruction of a cavity photon
leading to the creation of an exciton or vice versa.

Numerical simulations are performed using an open source quantum optics
toolbox~\cite{TanJOB99}. We calculate the two-time covariance function
$\langle \mathbf{a^{\dag}}(t+\tau), \mathbf{a}(t)\rangle$ in the steady state
limit using quantum regression theory.  The power spectrum is obtained by
taking the Fourier transform of the covariance function. We set the cavity
decay rate and the quantum dot-cavity coupling strength to the measured
values of $\kappa/2 \pi=36$~GHz and $g/2 \pi=15.3$~GHz, respectively. The
spontaneous emission rate is $\gamma/2 \pi=0.16$~GHz and the pure dephasing
rate is $\gamma_d/2 \pi=1$~GHz~\cite{FaveroPRB07}.

The sideband linewidth is calculated by fitting the calculated power spectrum
with a Lorentzian at each Rabi frequency. We determine $\gamma_{ph}^{a^{\dag}
\sigma_-}$ and $\gamma_{ph}^{a \sigma_+}$ in this simulation by treating them
as fitting parameters and performing a linear least squares optimization.
From the fit we obtain $\gamma_{ph}^{a^{\dag} \sigma_-}/2\pi =
0.19~(\pm0.03)$~GHz and $\gamma_{ph}^{a \sigma_+}/2\pi = 0.28~(\pm0.05)$~GHz.
We note that these phonon dephasing rates are consistent with the previously
predicted values~\cite{HughPRX11,RoyArx12}. The solid line in Fig.~3b shows
the theoretically calculated linewidths of the lower energy sideband as a
function of $|\Omega/2\pi|^2$ for $\Delta_{cx}/2\pi=42$~GHz. The
simulations exhibit good agreement with the measurement results and predict
anomalous broadening at the same excitation power. Fig.~3c plots the same
simulations for $\Delta_{cx}/2\pi=85$~GHz. Here we obtain phonon
coupling rates of $\gamma_{ph}^{a^{\dag} \sigma_-}/2\pi = 0.17~(\pm0.02)$~GHz
and $\gamma_{ph}^{a \sigma_+}/2\pi = 0.37~(\pm0.04)$~GHz. These values are
different from those obtained in Fig.~3b because the phonon dephasing rates
depend on $\Delta_{cx}$~\cite{HughPRX11}.

To explain the cause of anomalous broadening, we consider the situation where
there is no phonon mediated energy transfer or pure dephasing by setting
$\gamma_{ph}^{a^{\dag} \sigma_-}=\gamma_{ph}^{a \sigma_+}=\gamma_d=0$. The
blue dotted line in Fig.~4a shows the results for this simulation where the
detuning is set to $\Delta_{cx}/2\pi=42$~GHz. In this case the only
interaction between the quantum dot and cavity is through the coherent
Jaynes-Cummings term in the Hamiltonian.  The sideband linewidth exhibits a
resonant behavior that peaks when the higher energy sideband is on-resonance
with the cavity. The green dashed line shows the situation where we have
included pure dephasing, which does not change the resonant behavior but
simply broadens the sideband linewidth independent of the excitation power.
The model consisting of only a unitary Jaynes-Cummings interaction and pure
dephasing exhibits poor agreement with the experimental results. When we add
the phonon mediated dephasing terms (black solid line) we obtain a
significantly better agreement with the measured data. Fig.~4b plots the same
simulations for a detuning of $\Delta_{cx}/2\pi=85$~GHz.  None of the
simulations exhibit an anomalous power broadening because we are always in
the small Rabi frequency regime. Still, without the phonon term the
simulations predict a much smaller power induced broadening than what is
observed experimentally.

The simulations provide a clearer picture for the mechanism of anomalous
linewidth broadening.  Away from resonance, phonon coupling is the dominant
broadening mechanism (as shown in Fig.~4b) and results in a monotonic
increase of the sideband linewidth as a function of pump power.  Near
resonance, however, the coherent Jaynes-Cummings term significantly
contributes to the sideband linewidth (Fig.~4a). In the small Rabi frequency
regime, both phonon mediated dephasing and coherent Jaynes-Cummings
interaction exhibit an increase in linewidth as a function of pump power and
therefore constructively add.  In the large Rabi frequency regime, the
Jaynes-Cummings term exhibits a decrease in the linewidth as a function of
pump power. Instead of adding constructively, this decrease now partially
cancels out the monotonic increase due to phonon coupling.  The interplay
between these two terms therefore leads to the anomalous linewidth broadening
behavior observed in the experimental measurements.

In conclusion, we demonstrate that the Mollow triplet sideband of a quantum
dot exhibits a nonlinear dependence of the sideband linewidth with the
excitation power as well as emission enhancement when the Rabi frequency is
comparable to the quantum dot-cavity detuning. Our experimental results agree
well with recent theoretical prediction and provide further insight into the
Mollow triplet emission properties in the strong coupling regime.  These
results could find important applications for single and heralded photon
sources with large frequency tunability~\cite{UlhaqNPhon12}. They could also
provide a direct pathway for achieving population inversion with a resonantly
driven single emitter~\cite{QuangPRA93}.

We would like to acknowledge S.~Hughes for helpful discussions. This work was
supported by the ARO MURI on Hybrid quantum interactions (grant
no.~W911NF09104), the Physics Frontier Center at the Joint Quantum Institute,
a DARPA Defense Science Office grant (grant no.~W31P4Q0910013), and the ONR
Applied Electromagnetics center.  E.~Waks would like to acknowledge support
from an NSF CAREER award (grant no.~ECCS.~0846494) and a DARPA Young Faculty
Award (grant no.~N660011114121).

\begin{figure}
\centering
        \includegraphics[width=12cm, clip=true]{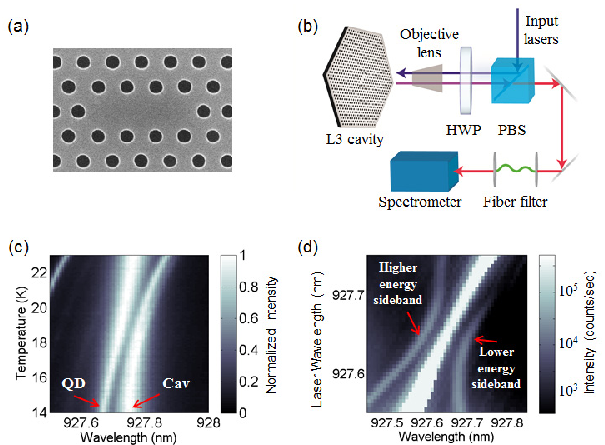}
    \caption[Fig1]{
(Color online) (a) Scanning electron microscope image of a fabricated
photonic crystal cavity. (b) Schematic of a measurement setup. HWP: half-wave
plate, PBS: polarizing beam splitter. (c) Photoluminescence spectra for a
coupled quantum dot-photonic crystal cavity device as a function of
temperature. The quantum dot and the cavity show an anti-crossing behavior,
indicating strong coupling. (d) Resonance fluorescence spectra when the
excitation laser is scanned across the quantum dot resonance at a pump power of
8.0~$\mu$W. Along with a strong laser signal, two Mollow triplet sidebands are observed.} \label{Fig1}
\end{figure}

\newpage

\begin{figure}
\centering
        \includegraphics[width=12cm, clip=true]{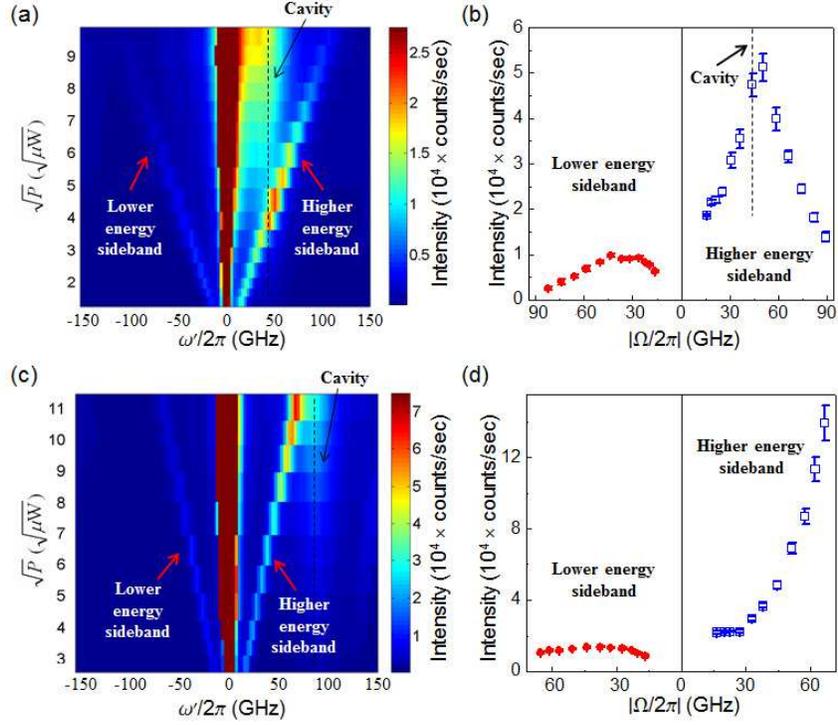}
    \caption[Fig2]{
(Color online) (a) Resonance fluorescence spectra of a quantum dot coupled to
a photonic crystal cavity as a function of $\sqrt{P}$ for
$\Delta_{cx}/2\pi=42$~GHz.  (b) Measured emission intensity of the
Mollow triplet sidebands as a function of $\Omega/2\pi$. (c) Same as panel a for $\Delta_{cx}/2\pi=85$~GHz. (d) Same
as panel b for $\Delta_{cx}/2\pi=85$~GHz.
\\
\\}
    \label{Fig2}
\end{figure}

\newpage

\begin{figure}
\centering
        \includegraphics[width=16cm, clip=true]{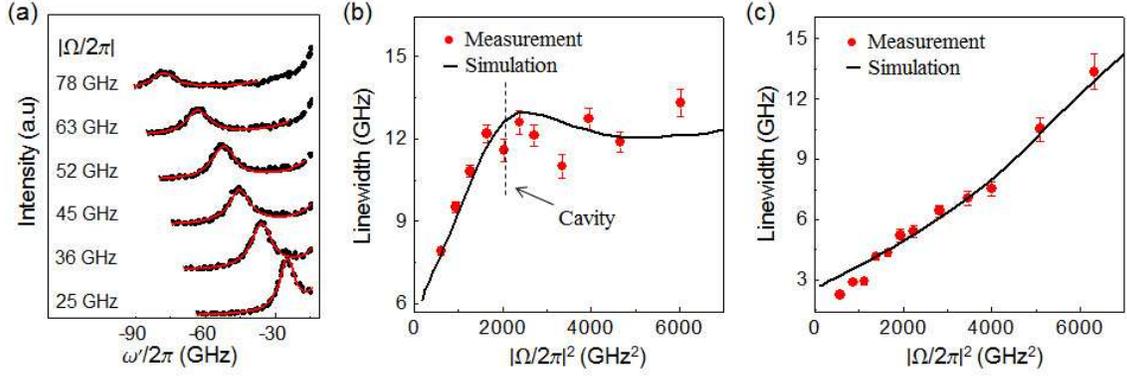}
    \caption[Fig3]{
(Color online) (a) High spectral resolution measurement (black circles) of
the lower energy sidebands using the Fabry-Perot filter for
$\Delta_{cx}/2\pi=42$~GHz. Red curves show the Lorentzian fit.
Measured and numerically calculated linewidths for (b) $\Delta_{cx}/2\pi=42$~GHz and
(c) $\Delta_{cx}/2\pi=85$~GHz.}
    \label{Fig3}
\end{figure}

\newpage

\begin{figure}
\centering
        \includegraphics[width=12cm, clip=true]{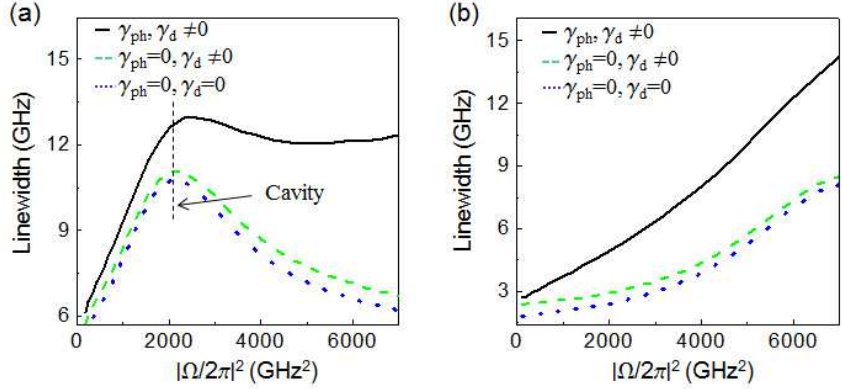}
    \caption[Fig4]{
(Color online) (a) Numerically calculated linewidths for
$\Delta_{cx}/2\pi=42$~GHz. Blue dotted line: calculated linewidth with
no phonon terms or pure dephasing, green dashed line: calculated linewidth with pure dephasing
but no phonon terms, black solid line: both phonon term and pure
dephasing included. (b) Same as panel a, but for $\Delta_{cx}/2\pi=85$~GHz.}
    \label{Fig4}
\end{figure}

\end{document}